# Electrical transport and ferromagnetism in $Ga_{1-x}Mn_xAs$ synthesized by ion implantation and pulsed-laser melting

M. A. Scarpulla,[1,2] R. Farshchi,[1,2] P. R. Stone,[1,2] R. V. Chopdekar,[3,1,2] K. M. Yu,[2] Y. Suzuki,[1,2] and O. D. Dubon[1,2]

[1]*Department of Materials Science and Engineering, University of California at Berkeley, Berkeley, California 94720, USA*
[2]*Lawrence Berkeley National Laboratory, Berkeley, California 94720, USA*
[3]*School of Applied and Engineering Physics, Cornell University, Ithaca, New York 14853, USA*

We present a detailed investigation of the magnetic and magnetotransport properties of thin films of ferromagnetic $Ga_{1-x}Mn_xAs$ synthesized using ion implantation and pulsed-laser melting (II-PLM). The field and temperature-dependent magnetization, magnetic anisotropy, temperature-dependent resistivity, magnetoresistance, and Hall effect of II-PLM $Ga_{1-x}Mn_xAs$ films have all of the characteristic signatures of the strong p-d interaction of holes and Mn ions observed in the dilute hole-mediated ferromagnetic phase. The ferromagnetic and electrical transport properties of II-PLM films correspond to the peak substitutional Mn concentration meaning that the non-uniform Mn depth distribution is unimportant in determining the film properties. Good quantitative agreement is found with films grown by low temperature molecular beam epitaxy (LT-MBE) and having the similar substitutional $Mn_{Ga}$ composition. Additionally, we demonstrate that II-PLM $Ga_{1-x}Mn_xAs$ films are free from interstitial $Mn_I$ because of the high temperature processing. At high Mn implantation doses the kinetics of solute redistribution during solidification alone determine the maximum resulting $Mn_{Ga}$ concentration. Uniaxial anisotropy between in-plane $[\bar{1}10]$ and $[110]$ directions is present in II-PLM Ga$_{1-x}$Mn$_x$As giving evidence for this being an intrinsic property of the carrier-mediated ferromagnetic phase.

---

INTRODUCTION

Investigations into spin-dependent phenomena and the prospects of applications such as the combination of non-volatile storage and processing of data on a single chip has driven the resurgence of research in diluted magnetic semiconductors over the last decade. Some of these material systems based on traditional semiconductors – particularly II-VI and III-V semiconductors alloyed with Mn – exhibit ferromagnetism mediated by delocalized holes [1-3]. $Ga_{1-x}Mn_xAs$ has emerged as the most studied and well-understood III-Mn-V material and is typically grown by low-temperature molecular beam epitaxy (LT-MBE) [3,4]. It has been demonstrated that $T_{CS}$ exceeding 170 K can be achieved in LT-MBE films after low-temperature post-growth annealing [5-7].

The use of ion implantation and pulsed-laser melting (II-PLM) to incorporate dopants in semiconductors was introduced in the late 1970s and remains an ongoing topic of research today. Most studies have focused on Si, where the incorporation of dopants like B, P, and As at concentrations in excess of $10^{21}$ /$cm^3$ has been realized [8-11]. Similar results have been reported in GaAs for traditional dopants such as Te [12-15] as well as for non-traditional dopants in Si such as Bi [16]. PLM has been used to synthesize both equilibrium phases such as transition metal silicides [17] as well as metastable phases like amorphous semiconductors [18]. This is due to the $10^9$-$10^{10}$ K/s quench rates achievable using nanosecond (ns) laser pulses; even faster cooling rates can be induced by femptosecond pulses [19]. The incorporation of impurities in semiconductors at concentrations exceeding maximal solubility limits without precipitation or the formation of second phases is due to this fast quenching.

We have previously demonstrated that the II-PLM process is a simple and versatile processing route for the formation of ferromagnetic semiconductors like $Ga_{1-x}Mn_xAs$ and $Ga_{1-x}Mn_xP$ [20-24] as well as for highly-mismatched semiconductor alloys [25-27]. In this work, we present a detailed investigation of the ferromagnetic and electrical transport properties of $Ga_{1-x}Mn_xAs$ films synthesized by II-PLM. We show that all of the general features of temperature and field dependencies of magnetization, sheet resistivity, and Hall resistivity agree with results reported for properly annealed $Ga_{1-x}Mn_xAs$ films synthesized using LT-MBE. Furthermore, good quantitative agreement is found between results from II-PLM films characterized by their maximum substitutional Mn concentration and from LT-MBE films of equivalent substitutional Mn concentration.

EXPERIMENTAL

Semi-insulating GaAs (001) wafers were implanted at 7° from the surface normal with 50 or 80 keV $Mn^+$ to doses of $1.5 \times 10^{16}$, $1.8 \times 10^{16}$, or $2.0 \times 10^{16}$ /cm$^2$. Table 1 summarizes the four types of samples discussed herein. 50 and 80 keV $^{55}Mn^+$ have ranges in GaAs of 32 and 49 nm respectively as calculated with SRIM [28]. Analysis by Rutherford backscattering spectrometry (RBS) and particle-induced X-ray emission (PIXE) [7] using 1.95 MeV $^4He^+$ revealed that a layer approximately 70-100 nm thick is amorphized and that Mn loss from sputtering during implantation is negligible. Samples measuring approximately 5 mm on a side were cleaved from implanted wafers producing <110> edges. Each sample was irradiated in air with a single pulse from a KrF excimer laser (λ = 248 nm) with duration ~32 ns, FWHM 23 ns, and peak intensity at 16 ns. The KrF laser pulses having fluence of 0.2 – 0.4 J/cm$^2$ pass through a crossed-cylindrical lens homogenizer which produces a very uniform spatial intensity distribution of ±5 % – thus the film properties are uniform across each sample. RBS/PIXE ion channeling analysis demonstrated that the fraction of Mn residing on substitutional sites (Mn$_{Ga}$) was typically 75-85 % depending on the Mn implanted dose and laser fluence. During PLM, the ion-implanted region of the film melts, solidifies epitaxially, and then cools to room temperature within a few hundred nanoseconds. As a result of this high-temperature processing (T$_{Melt}$=1511 K for GaAs), films produced using II-PLM are free from interstitial Mn$_I$ and the post-growth annealing typically carried out on LT-MBE films is unnecessary to achieve high T$_C$ [20,22]. Etching in concentrated HCl for 5-20 minutes was used to remove excess Mn from the surface that was present in Ga droplets and in surface oxides. It was verified that this etching did not affect the ferromagnetic or electrical transport properties of the films. Anisotropic etching using a solution of KI:I:H$_2$SO$_4$ was used to verify the crystalline orientation of the sample after the measurements of magnetic anisotropy [29].

Secondary ion mass spectrometry (SIMS) was performed using a Cs$^+$ beam and detection of CsMn$^+$ (188 amu) ions. Film magnetization was measured with a SQUID magnetometer using a field of 50 Oe for temperature-dependent measurements. The total amount of Mn (Mn$_{Ga}$ and non-commensurate Mn) was determined from the sample area and PIXE measurements. T$_C$ estimations were made by extrapolating the steepest portion of the temperature-dependent data to zero magnetization and are reported with an uncertainty of 2 K which is typical of the sample-to-sample variation for a given set of processing conditions. Magnetotransport measurements were made in the van der Pauw geometry with the field applied perpendicular to the sample plane using cold-pressed indium contacts. Due to the large p-type doping, these contacts are Ohmic as evaluated by I-V curves taken at each temperature and field combination. Field symmetrization was used to remove even-parity contributions from the Hall data and odd-parity contributions from the magnetoresistance data. Additional magnetoresistance measurements were also carried out with the field in-plane to investigate isotropic and magnetization-direction-dependent contributions.

RESULTS & DISCUSSION

Figure 1 displays a SIMS profile of a sample of Type A. The Mn distribution extends to approximately 120 nm, peaks at x=0.051 near 25 nm, and has a FWHM of ~60 nm. The shape of this Mn depth distribution is typical for all of the samples discussed herein with differences being in the total retained dose and small changes in the center and width of the distribution due to implanted ion energy. RBS / PIXE measurements indicate that $7.2 \times 10^{15}$ /cm$^2$ Mn (48 %) was retained after irradiation and HCl etching and that 75 % of this retained Mn exists as substitutional Mn$_{Ga}$. This implies that the peak substitutional Mn$_{Ga}$ concentration is x=0.038 assuming no dependence of Mn substitutionality on composition. Samples such as this one exhibit $T_C$ of 100 K, which is slightly higher than that for annealed LT-MBE samples having x=0.034 or 0.045 [30]. The saturation magnetization is 3.2 ± 0.3 $\mu_B$ per total Mn, corresponding to 4.3 ± 0.4 $\mu_B$ per substitutional Mn$_{Ga}$, which is in excellent agreement with the best measurements from LT-MBE grown $Ga_{1-x}Mn_xAs$ [30,31]. For samples of Type B, a peak total Mn concentration near x=0.10 and a FWHM of 50 nm were measured.

Figure 2 displays the in-plane magnetization along the [$\bar{1}$10] in-plane orientation for $Ga_{1-x}Mn_xAs$ samples synthesized using slightly different implantation and laser melting conditions. The solid lines are data from a sample of Type B; this particular film has a $T_C$ of 132 K and displays square hysteresis loops both at 5 K and 100 K (inset). The dashed lines are data from a sample of Type C having $T_C$ of 94 K. Samples of Type D have $T_C$s of 130 K similar to samples of Type B [22]. The details of the ferromagnetic hysteresis loops at 5 K and low fields appear to be somewhat insensitive to the II-PLM processing parameters in this range of ion doses.

These data suggest that the kinetics of solute redistribution during solidification determine the maximum resulting $Mn_{Ga}$ concentration (as opposed to the implanted dose) for ion implantation conditions resulting in high solute concentrations. Ion implantation of $^{55}Mn^+$ into GaAs at energies of 50-80 keV (30-50 nm range, respectively) and doses approaching $2 \times 10^{16}$ /cm$^2$ results in maximum Mn concentrations near x=0.15. The kinetic limitation on the maximum $Mn_{Ga}$ concentration comes about through the balance between solidification velocity and solute diffusion away from the liquid-solid interface [16]. The solidification velocity is determined primarily by the laser pulse duration and thermal diffusion coefficient of the sample, thus higher Mn concentrations could be achieved by using a laser with a shorter pulse width. We have also demonstrated that at lower implanted Mn concentrations (below approximately $1 \times 10^{16}$ /cm$^2$ $^{55}Mn^+$ at 80 keV) the implanted dose determines the resulting maximum MnGa concentration and thus film $T_C$.

The magnetic anisotropy of epitaxial $Ga_{1-x}Mn_xAs$ films is complex and is affected by strain, carrier concentration, and temperature. The Zener model has been applied to explain these effects including a peculiar in-plane anisotropy term breaking the symmetry between the $[\bar{1}10]$ and $[110]$ directions [32-35]. This symmetry breaking is also present in $Ga_{1-x}Mn_xAs$ synthesized by II-PLM, adding further evidence that it is indeed an intrinsic property of the hole-mediated ferromagnetic phase and not due to LT-MBE processing effects. Figure 3 presents hysteresis loops measured at 5 K along the $[\bar{1}10]$, $[110]$, $[100]$, and out-of-plane $[001]$ directions from a sample of Type A. It is seen that the $[\bar{1}10]$ direction is the easiest direction as it has the highest remanent magnetization. The coercivity along the $[\bar{1}10]$ direction is 150 Oe, while along the $[110]$ direction the coercivity is 95 Oe. The out-of-plane $[001]$ direction is the hardest direction and has coercivity of 200 Oe. The $[100]$ and $[010]$ in-plane directions are equivalent and identical in their magnetization as functions of temperature and field; they have coercivity near 115 Oe. These magnetic anisotropy characteristics including the inequivalence of in-plane $<110>$ directions are similar to those present in $Ga_{1-x}Mn_xAs$ layers epitaxially grown on GaAs using LT-MBE. However at low-temperature in LT-MBE samples the easy axes are typically close to $<100>$ while in II-PLM samples the easiest axis is $[\bar{1}10]$ followed closely by the $[100]$ and $[010]$ directions and then the $[110]$ direction [33,36]. This suggests that the ratio between the in-plane uniaxial anisotropy term and in-plane cubic term is stronger in II-PLM films than in LT-MBE films. The fact that magnetization along the $[\bar{1}10]$ axis is easier than along $[110]$ demonstrates that the in-plane uniaxial anisotropy in II-PLM films is equivalent to that in post-growth annealed

LT-MBE films [37] which is consistent with the absence of $Mn_I$ due to the high-temperature processing.

Figure 4 presents the temperature dependent sheet resistivities for two $Ga_{1-x}Mn_xAs$ samples of Types B and C prepared by II-PLM under conditions identical to those in Fig. 2 as well as for a sample of Type A. The resistivity at 300 K of the Type B film is estimated at ~2.5x10$^{-3}$ Ω·cm using the characteristic film thickness of 50 nm. This is nearly identical to the resistivity of 1.9x10$^{-3}$ Ω·cm reported for films having x=0.06 and $T_C$=140 K grown by LT-MBE and annealed to remove interstitial Mn [38]. The peaked behavior associated with $T_C$ is characteristic of spin-disorder scattering in disordered ferromagnets like $Ga_{1-x}Mn_xAs$ which are near the metal insulator transition [3,39-41]. The fact that the width of this resistivity peak is comparable to widths measured on LT-MBE $Ga_{1-x}Mn_xAs$ films rather than being broadened due to the non-uniform Mn depth distribution suggests that electrical transport is dominated by a portion of the films having relatively uniform Mn concentration. As $T_C$ and electrical conductivity are known to scale in $Ga_{1-x}Mn_xAs$ with $Mn_{Ga}$ concentration [1,3] the correspondence of the peak and film $T_C$ indicates that the portion of the films dominating both ferromagnetic and transport properties is the region having the highest $Mn_{Ga}$ concentration. Thus both the $T_C$ [20] and electrical transport behaviors can be correlated with the peak $Mn_{Ga}$ concentration for II-PLM films when this concentration can be established by the appropriate techniques such as SIMS and RBS/PIXE or by electrochemical capacitance-voltage profiling [42].

Figure 5 displays the (a) field and (b) temperature dependencies of the magnetoresistance measured with the field perpendicular to the plane ($MR_\perp$) for a sample of Type D having a $T_C$ of 132 K. Herein, MR is defined as the normalized difference between the sheet resistance in a magnetic field and with no field – MR={ R(H)-R(0) }/R(0). As for LT-MBE grown films, an isotropic negative component to $MR_\perp$ dominates for fields larger than ~5 kOe at all temperatures. A positive component varying with the angles θ between the magnetization vector $\vec{M}$ and the principal crystallographic axes and ϕ between the current vector $\vec{J}$ and $\vec{M}$ is also present at lower fields in the ferromagnetic regime [43-45]. This (θ,ϕ)-dependent component arises through the spin-orbit interaction and is closely related to the phenomenon of anisotropic magnetoresistance (AMR). The isotropic component does not saturate for fields up to 50 kOe (5 T) at any temperature down to 5K and reaches a magnitude of just over 6 % at 50 kOe and close to $T_C$. As displayed in Fig. 5(b), the negative isotropic MR has its largest magnitude at $T_C$ due to the divergent magnetic susceptibility [45]. Below $T_C$, the field dependence of the resistivity and $MR_\perp$ is cusp-like showing upward concavity while above $T_C$ the data are downward concave, as is demonstrated by the data at 160 K. Early results indicating orders-of-magnitude changes in resistivity in LT-MBE grown $Ga_{1-x}Mn_xAs$ appear to be due to compensation by interstitial $Mn_I$ [46]. The data in Fig. 5(b) suggests that the negative MR in II-PLM $Ga_{1-x}Mn_xAs$ films may be somewhat stronger than in post-growth annealed LT-MBE films despite the similar sheet resistivities [45,47].

In order to differentiate the isotropic and (θ,ϕ)-dependent contributions to MR arising from spin disorder scattering and spin-orbit interactions (respectively), resistivity measurements were made with the field aligned along an in-plane <110> direction as well as the out-of-plane [001] direction. Note that in this experiment the out-of-plane uniaxial anisotropy is primarily probed because it dominates the in-plane anisotropy components. We define ΔMR as the in-plane magnetoresistance (MR$_{IP}$) minus the out-of-plane (MR$_\perp$) magnetoresistance. As demonstrated in Fig. 6 for a sample of Type B, for T<T$_C$ a positive (θ,ϕ)-dependent component is added to the isotropic negative MR for $\vec{H}$ transverse to $\vec{J}$, i.e. for $\vec{H}$ out-of-plane. As in LT-MBE films [45], II-PLM Ga$_{1-x}$Mn$_x$As exhibits a positive (θ,ϕ)-dependent contribution at low to intermediate fields. It should be noted that this (θ,ϕ)-dependent MR extends well beyond the hysteretic region (small fields) of the sample magnetization shown in Fig. 2(b) indicating that its origin must lie in rotations of the magnetic moment against the magnetic anisotropies and not in domain wall motion. For $|\vec{H}|$ greater than ~10 kOe, this positive MR component is absent and only the isotropic negative MR is present. This difference in behavior is due to the magnetocrystalline anisotropy of the film; the easy magnetization axes of these II-PLM Ga$_{1-x}$Mn$_x$As lie in the plane, while the out-of-plane direction is the hardest axis. For $\vec{H}$ out-of-plane and $|\vec{H}| < H_a$ (where H$_a$ is the out-of-plane anisotropy field) as $|\vec{H}|$ increases the magnetization rotates out of plane until it aligns with the applied field at H$_a$. When ϕ=0, the (θ,ϕ)-dependent effect disappears; thus H$_a$ can be estimated for this film to be approximately 10 kOe (1 T) as only the isotropic component is present at larger fields.

Figure 7 (a) displays the field dependencies of the Hall resistance for an II-PLM Ga$_{1-x}$Mn$_x$As sample of Type C having T$_C$ of 95 K. The determination of the carrier density in Ga$_{1-x}$Mn$_x$As using the Hall effect without correction is known to underestimate the true carrier concentration due to the influence of the anomalous Hall component which is still significant even at temperatures up to 3·T$_C$ [48]. With this caveat, taking the slope of the Hall data for this II-PLM sample at 300 K indicates a carrier concentration of $5.3 \times 10^{20}$ /cm$^3$ again assuming a thickness of 50 nm. It is immediately apparent that the anomalous Hall component dominates the Hall resistance even above $T_C$ and that it reflects the switching of the ferromagnetic moment. These are traits of the dilute hole-mediated ferromagnetic phase in Ga$_{1-x}$Mn$_x$As and are necessary but not sufficient characteristics to prove the presence of this phase [49]. Again assuming a characteristic thickness of 50 nm, the data at 10 K indicate a Hall resistivity of $8.0 \times 10^{-5}$ Ω·cm. This is in excellent quantitative agreement with the Hall resistivity of $9.0 \times 10^{-5}$ Ω·cm reported for annealed LT-MBE samples having x=0.06, a similar thickness of 45 nm, similar resistivities of $1.4$-$5.0 \times 10^{-3}$ Ω·cm, and reasonably similar $T_C$s near 125 K [45]. The temperature dependence of the Hall resistance shown in Fig. 7(b) is also similar to that reported for LT-MBE Ga$_{1-x}$Mn$_x$As. In both cases it has the same rapid increase as the sample is cooled through $T_C$ due to the anomalous Hall effect (which is proportional to the magnetization) followed by a slight decrease below $T_C$ due to the decreasing resistivity (Fig. 4) as the magnetization approaches saturation.

In summary, we have investigated the electrical transport and ferromagnetic properties of Ga$_{1-x}$Mn$_x$As synthesized using II-PLM and found good quantitative agreement with properties reported for films grown using LT-MBE. This is an important finding because it demonstrates that reliable quantitative results can be obtained using II-PLM and to date films of Ga$_{1-x}$Mn$_x$P displaying the same carrier-mediated ferromagnetic phase have only been synthesized using II-PLM [21]. Theories of the carrier mediated ferromagnetic phase in dilute ferromagnetic semiconductors must be able to account for the ferromagnetism in insulting $Ga_{1-x}Mn_xP$ and other systems such as low-doped $Ga_{1-x}Mn_xAs$ where carriers reside in an impurity band [50,51].

As the electrical and magnetic properties of II-PLM $Ga_{1-x}Mn_xAs$ films may be correlated with the peak $Mn_{Ga}$ concentration, quantitative comparisons to films grown by LT-MBE may be made. II-PLM $Ga_{1-x}Mn_xAs$ films display the magnetic anisotropy, resistivity, magnetoresistance, and anomalous Hall effect signatures characteristic of the strong *p-d* interaction of holes and Mn ions observed in the dilute hole-mediated ferromagnetic phase observed in LT-MBE $Ga_{1-x}Mn_xAs$. Good quantitative agreement is found with films grown by low temperature molecular beam epitaxy (LT-MBE) and having the similar substitutional $Mn_{Ga}$ composition.

Thanks to T. Mates for SIMS measurements and S. Boettcher for help with etching. This work is supported by the Director, Office of Science, Office of Basic Energy Sciences, Division of Materials Sciences and Engineering, of the U.S. Department of Energy under Contract No. DE-AC02-05CH11231. MAS acknowledges support from an NSF Graduate Research Fellowship. PRS acknowledges support from a NDSEG Fellowship.

| Sample Type | Mn$^+$ Dose (/cm$^2$) | Mn$^+$ Energy (keV) | Laser Fluence (J/cm$^2$) | T$_C$ (K) |
|---|---|---|---|---|
| A | 1.5x10$^{16}$ | 50 | 0.30 | 100±2 |
| B | 2.0x10$^{16}$ | 50 | 0.20 | 131±2 |
| C | 1.8x10$^{16}$ | 80 | 0.40 | 94±2 |
| D | 1.8x10$^{16}$ | 80 | 0.2 | 131±2 |

TABLE 1 - Scarpulla *et al.*

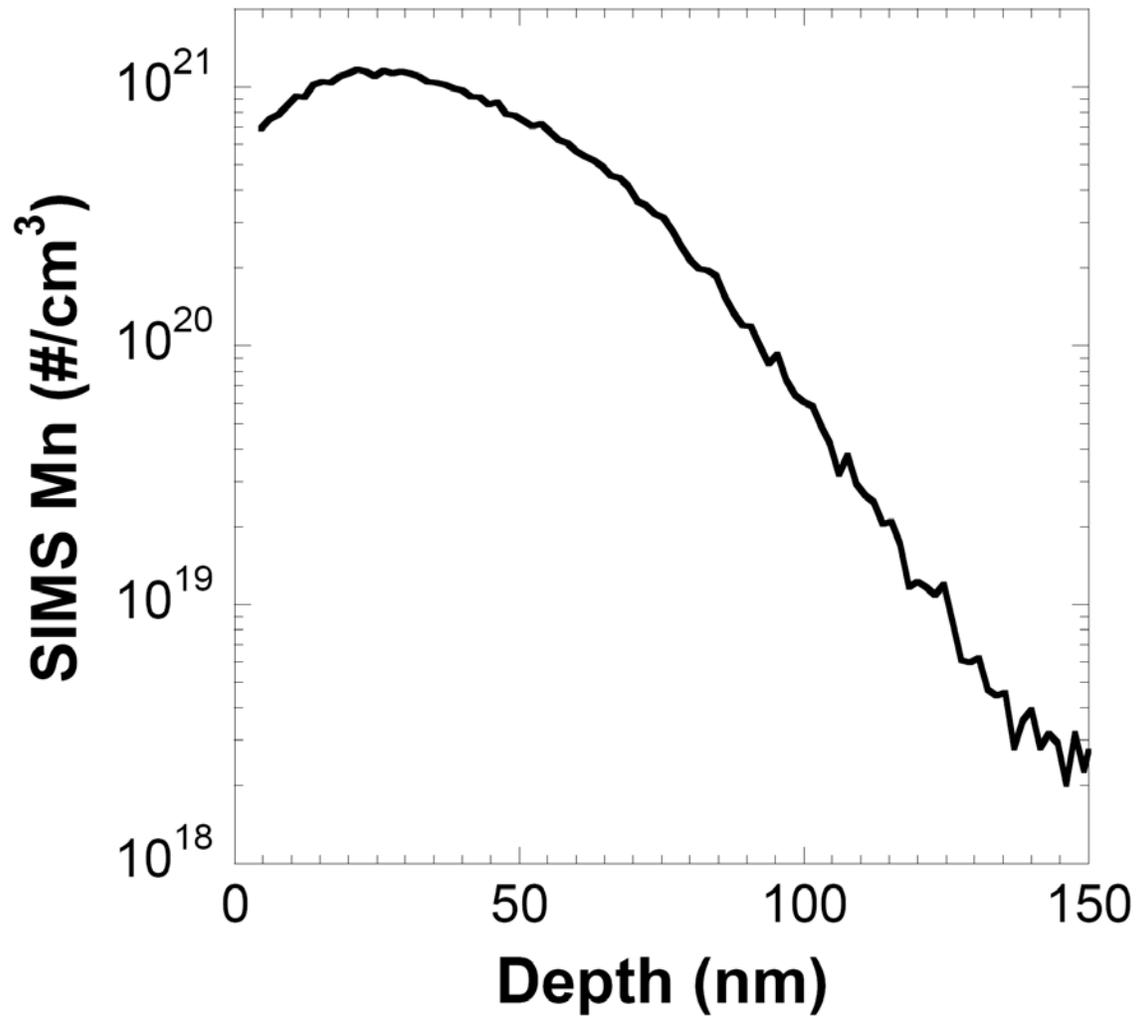

FIGURE 1 - Scarpulla *et al*.

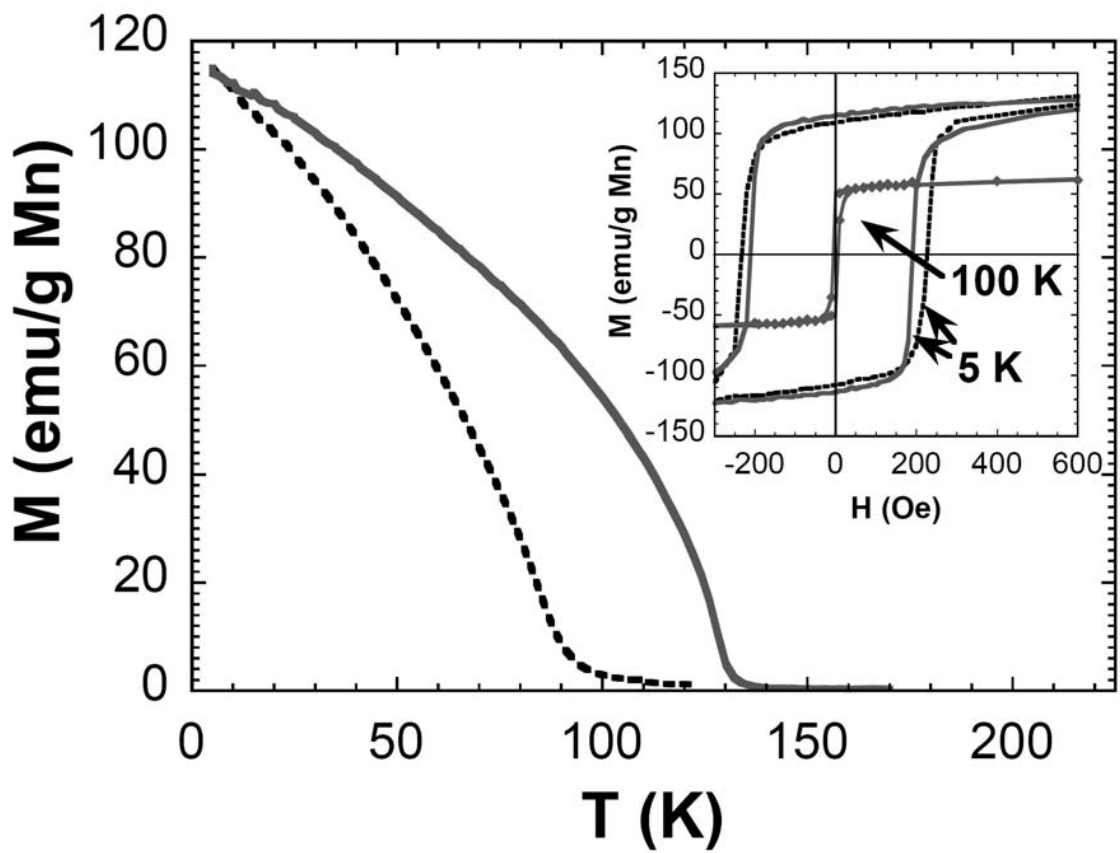

FIGURE 2 - Scarpulla *et al.*

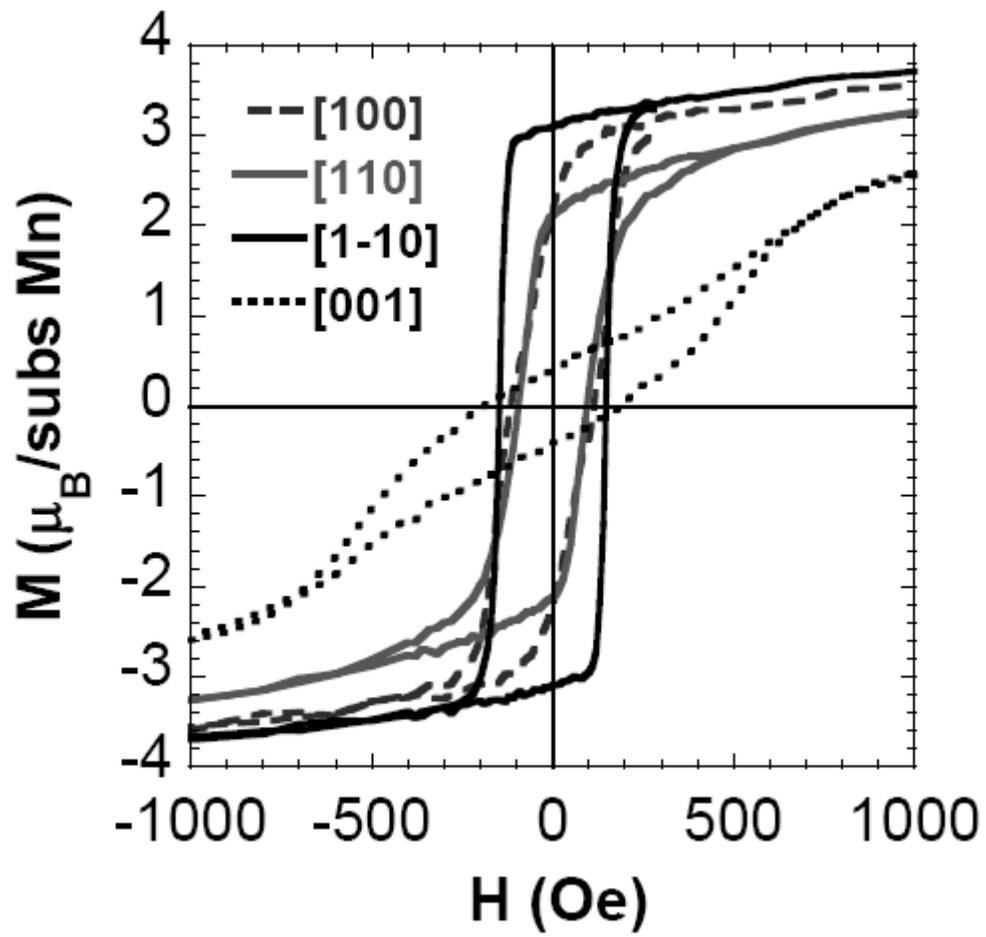

FIGURE 3 - Scarpulla *et al.*

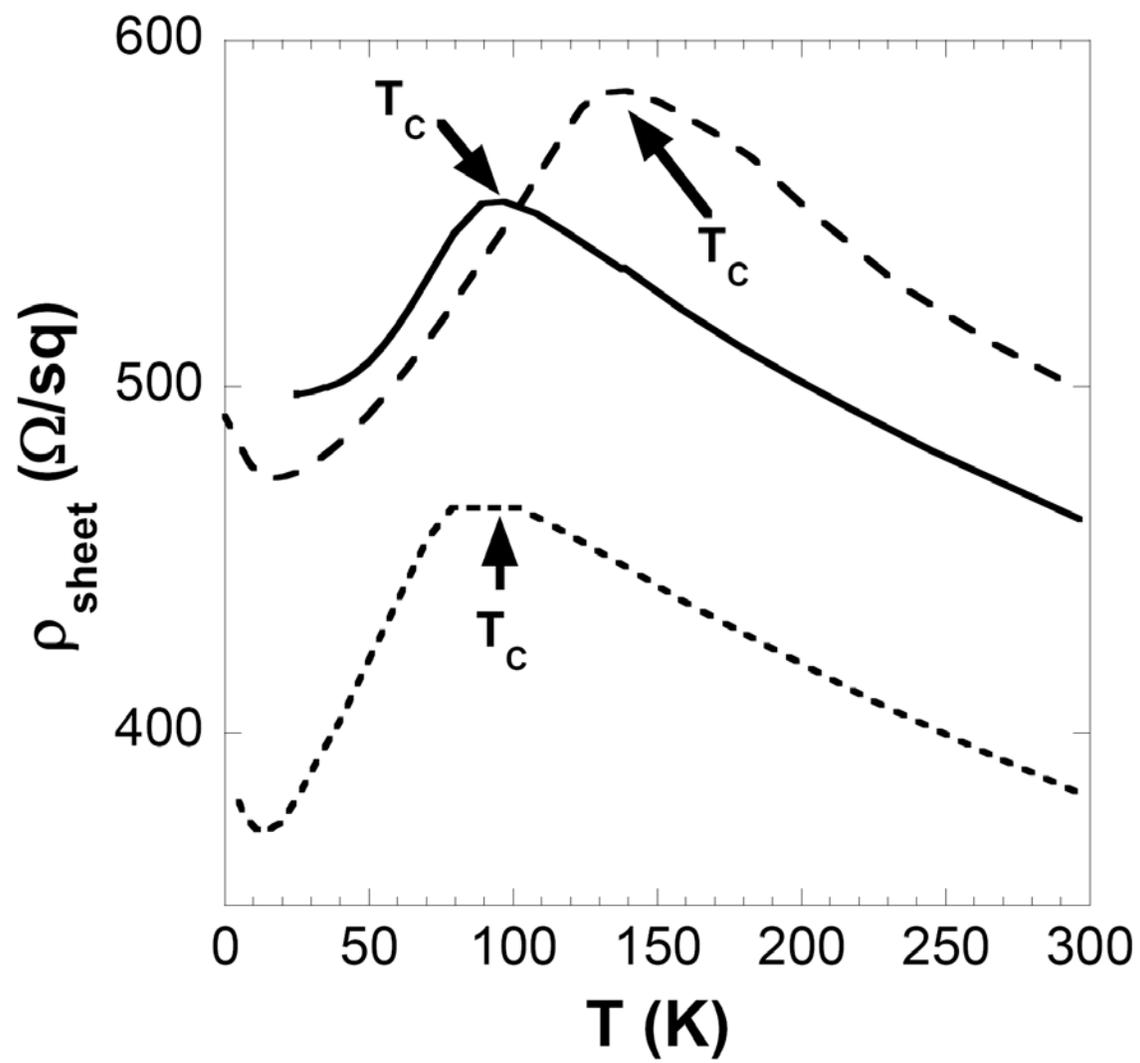

FIGURE 4 - Scarpulla *et al.*

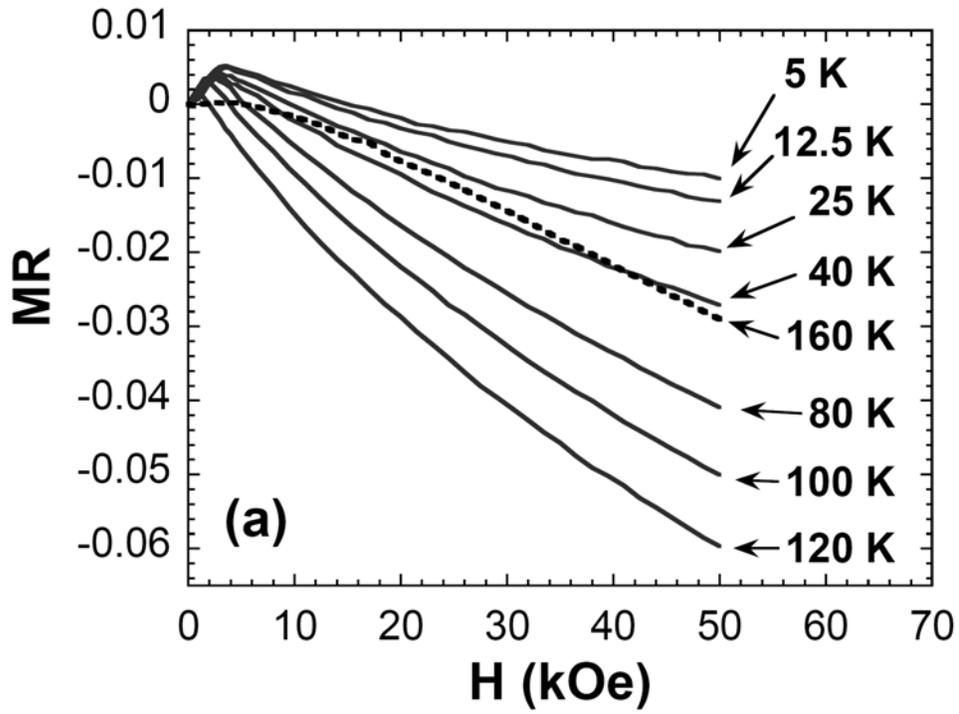
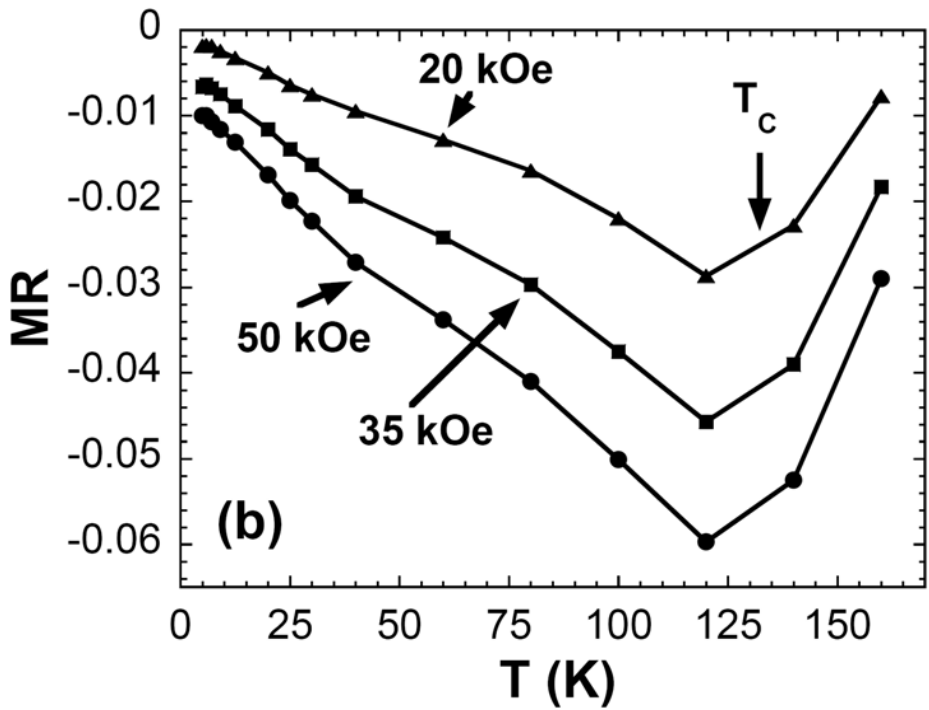

FIGURE 5 - Scarpulla *et al.*

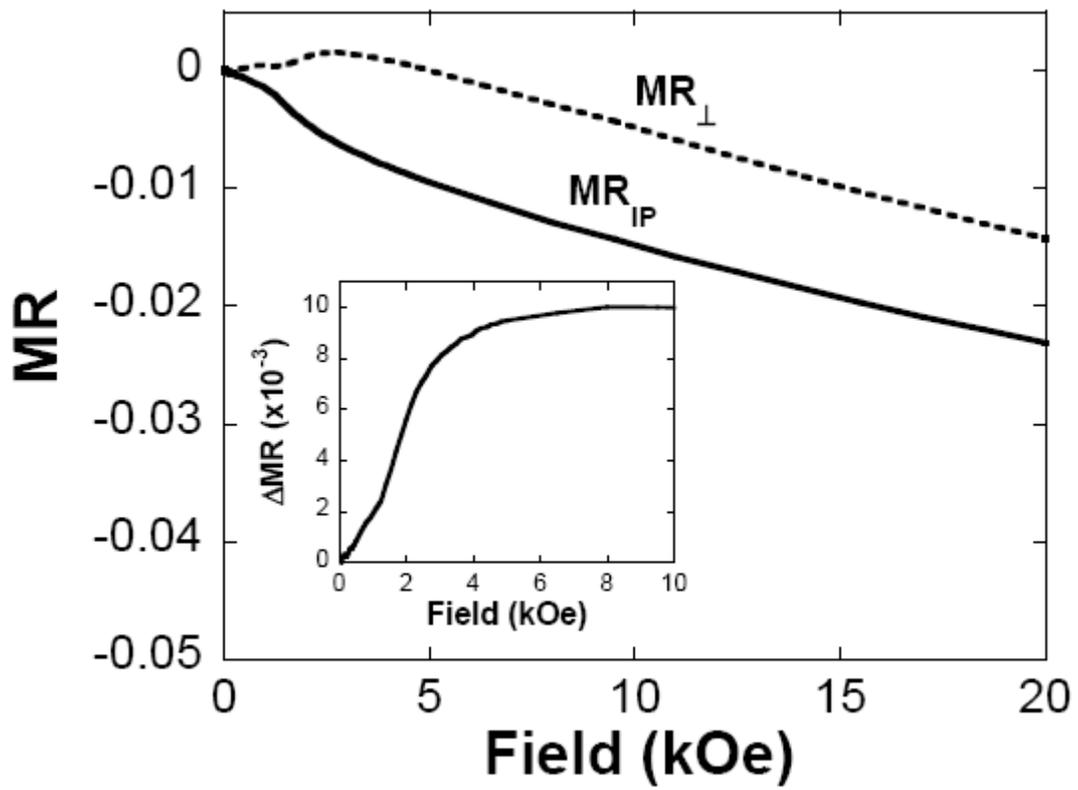

FIGURE 6 - Scarpulla *et al.*

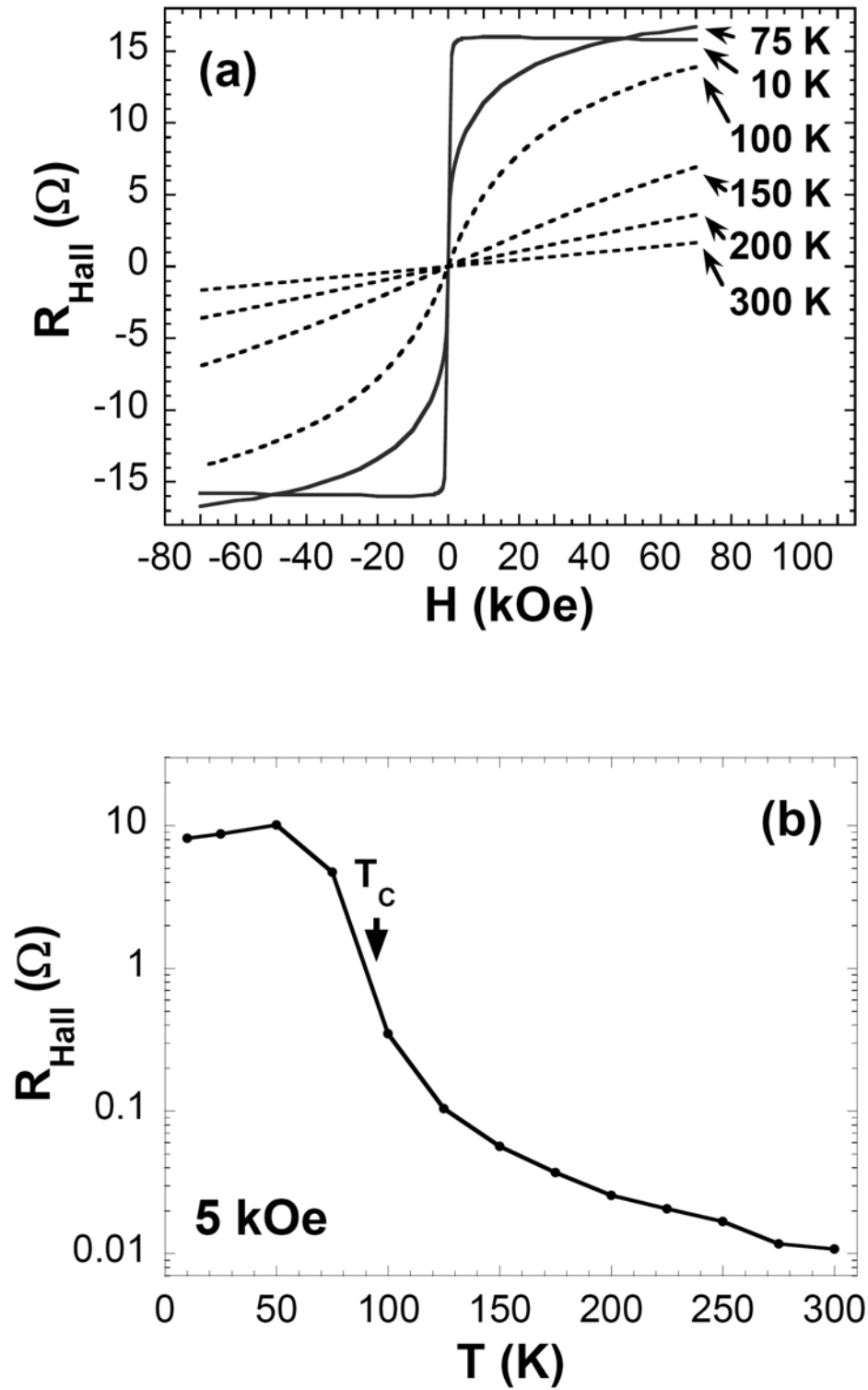

FIGURE 7 - Scarpulla *et al.*

FIGURE & TABLE CAPTIONS

**Table 1** – Summary of ion implantation and laser melting conditions used for samples discussed in the text. The 2 K uncertainty in $T_C$ is caused by sample-to-sample variation.

**Figure 1** – SIMS depth profile of the Mn concentration in a sample of Type A. Such samples have peak x=0.051, 75 % of Mn substitutional ($Mn_{Ga}$), $T_C$=100 K, and $M_{sat}$=4.3 ± 0.4 $\mu_B/Mn_{Ga}$.

**Figure 2** – [main] <110> in-plane temperature dependent magnetization for a sample of Type C (dashed) and a sample of Type B (solid) having $T_C$s of 95 and 132 K, respectively. [inset] Field-dependent magnetization for the same two samples in the same orientation. The M vs. H data shown by a solid line and diamonds is from the Type B sample in the same orientation but at 100 K.

**Figure 3** – Magnetization measured as a function of field along major crystallographic directions for a sample of Type A. The easiest axis is the $[\bar{1}10]$ in-plane direction, while the [001] out-of-plane direction is the hardest axis. The in-plane uniaxial component is evidenced by the inequivalent hysteresis loops for the $[\bar{1}10]$ and [110] directions.

**Figure 4** – Temperature dependence of sheet resistivity for a sample of Type A (solid), a sample of Type B (dashed), and a sample of Type C (dotted). The peaked behavior of the resistivity associated with $T_C$ is a feature of the hole-mediated ferromagnetic phase in $Ga_{1-x}Mn_xAs$.

**Figure 5** – (a) Field and (b) temperature dependencies of MR for a sample of Type D. In (a), data above the $T_C$ of 130 K is shown as dashed lines, while solid lines are used for data below $T_C$. These MR behaviors are very similar to what has been reported for LT-MBE films.

**Figure 6** – [main] MR at 5 K for field applied in-plane ($MR_{IP}$) and perpendicularly ($MR_\perp$) as functions of applied field for a sample of Type B having a $T_C$ of 135 K. [inset] $\Delta MR$ at 5 K. The out-of-plane anisotropy field is estimated from this data to be 10 kOe.

**Figure 7** – (a) Field and (b) temperature dependencies of the Hall resistance measured in the van der Pauw geometry for a sample of Type C. In (a), data above the $T_C$ of 95 K is shown as dashed lines, while solid lines are used for data below $T_C$. The strong anomalous Hall signal reflecting the sample magnetization is strong evidence for the hole-mediated origin of ferromagnetism.